\documentstyle[prb,aps,eqsecnum]{revtex}
\begin{document}


\title{Coexistence of $0$ and $\pi$ states  in
Josephson junctions\footnote{Shorter version, without Figs. 6,7,8, is submitted
to Phys. Rev. B, October 19. 2000.}}
\author{Z. Radovi\'c}
\address{Faculty of Physics, University of Belgrade, P.O. Box 368, 11001 Belgrade, Yugoslavia}
\author{L. Dobrosavljevi\'{c}--Gruji\'{c}}
\address{Institute of Physics, P.O. Box 57, 11080 Belgrade, Yugoslavia}
\author{B. Vuji\v{c}i\'c}
\address{Faculty of Science, University of Montenegro, P.O. Box 211, 81000 Podgorica, Yugoslavia}

\maketitle

\begin{abstract}
New modes of  the magnetic flux penetration  in SQUIDs with
ferromagnetic junctions are predicted theoretically.
Characteristic multinode anharmonicity of the current-phase
relation implies the coexistence of stable and metastable $0$ and $\pi$
states for a large interval of strength of the ferromagnetic barrier
influence at low temperature. As a consequence, the coexistence of
integer and half-integer fluxoid configurations appears in both rf and
dc SQUIDs, and
generates two flux jumps per one external flux quantum. In dc SQUIDs,
new flux jumps are manifested  as two   dips in the critical current
dependence on the external magnetic flux.
\end{abstract}


\section{Introduction}

A number of possibilities to obtain negative coupling between two
superconductors in a Josephson junction have been
proposed.\cite{bul,spivak,sig} The negative sign of the Josephson
current corresponds to the coupling energy minimum at the  phase
difference $\pi$  between the superconducting electrodes. In
superconducting rings containing an odd number of such $\pi$ shifts, the
flow of spontaneous supercurrent may lead to anomalous flux
quantization. This has been confirmed in experiments with high-$T_{\rm
c}$ superconductors, where the negative coupling is due to  $d$-wave
symmetry of the superconducting order parameter.\cite{har} Also,
reversing the direction of the supercurrent appears to be observed in
out-of-equilibrium mesoscopic superconducting structures.\cite{base}
Recently, a metastable  $\pi$   state has been observed in $^3$He weak
links.\cite{mar}

The possibility of $\pi$ coupling through a barrier containing
paramagnetic impurities has been proposed first by Bulaevskii et
al.,\cite{bul} and in superconductor/ferromagnet/superconductor (S/F/S)
metallic weak links by Buzdin at al.\cite{bbp} Subsequently, the
progress in fabrication of S/F  multilayers has motivated considerable
experimental\cite{wong,strunk,koor,jiang,muhge,ver,mer,obi} and
theoretical\cite{91,bk,dem,kub,tagirov,andr,vesna} investigations of
$\pi$ coupling. Theory of proximity coupled S/F superlattices in the
dirty limit,\cite{91,bk,dem,kub} and of atomic-scale S/F superlattices
with tunneling between S and F layers,\cite{andr,vesna} is well
established.  A clear evidence of $\pi$ coupling, which has been sought
in several S/F systems from the oscillations of the critical
temperature, predicted in Refs. \onlinecite{91} and \onlinecite{bk}, now
appears to be established unambiguously.\cite{obi,lazar} Recently,
another evidence of $\pi$ state is found in the nonmonotonic variation
of the critical current in S/F/S junctions.\cite{olivier,rjaz}

The modified Andreev reflection at S/F interfaces  has been explored
theoretically, both for $s$-wave\cite{jong} and $d$-wave
superconductors,\cite{zhu} and experimentally.\cite{upa,vasko} The
Josephson current in $s$-wave superconductor/ferromagnetic
insulator/supercon\-ductor (S/FI/S) junctions has been calculated in
several papers,\cite{kup,deweert,tanC} whereas more recent works deal
with $d$-wave S/F/S and F/I/S junctions,\cite{rad,kt99} and with the
general case of unconventional superconductor junctions with
ferromagnetic barrier.\cite{tanJ}

At low temperature the Andreev reflection is dominant, and the
 ferromagnetic barrier may induce a characteristic anharmonic
 current-phase relation with new nodes in $I(\phi)$.\cite{tanC,rad,tanJ}
 The similar type of anharmonicity may appear in  $d$-wave pinhole
 junctions,\cite{fog} or in  S/I/S junctions,\cite{tan97} as found
 experimentally in the grain-boundary Josephson junctions in
 high-$T_{\rm c}$ superconductors.\cite{ilj}

In this paper we present new consequences of the modified Andreev
reflection in S/F/S junctions. We show that coexisting stable and
metastable $0$ and $\pi$ states appear in the vicinity of a critical
strength of the ferromagnetic barrier influence, Sec. II.  This coexistence
leads to new modes of the magnetic  flux penetration in superconducting loops
interrupted by a single ferromagnetic junction (rf SQUID), or two
junctions (dc  SQUID), Sec. III and IV, respectively. Section V
contains a brief summary and discussion.

\section{Current-phase relation}

We  consider  an  S/F/S junction with a   ferromagnetic metal barrier of
thickness $d$ and  constant exchange energy $h$.  Both superconducting
and ferromagnetic metals are assumed clean, with same dispersion
relations and with same Fermi energies. The pair potential variation is
taken as a step function:  $\Delta =0$ in the barrier,  and
$\Delta=\Delta_0\exp({\pm i\phi/2})$  in superconducting electrodes, where
$\phi$ is the phase difference at the junction.

The Josephson current density is calculated within the
quasiclassical theory of superconductivity,\cite{bbp}
\begin{equation}
{\bf j}({\bf r}) = -2ie\pi N(0)k_BT\sum_{\omega_n}
\left\langle {\bf v}_0 {g_\uparrow
+g_\downarrow\over2}\right\rangle ,
\label{current}
\end{equation}
where ${\bf v}_0$ is the Fermi velocity, $\langle\cdots \rangle$
denotes the angular averaging over the Fermi surface, and $N(0)$ is
the density of states at the Fermi level.
The quasiclassical normal Green's functions $g_\sigma ({\bf r, {\bf
v}_0},\omega_n)$,   where $\sigma=\uparrow,\downarrow$, and $\hbar
\omega_n = \pi k_BT(2n+1)$, are calculated from the Eilenberger
equations in  the previous
paper.\cite{rad}

For $s$-wave pairing in superconducting  electrodes and
transparent interfaces between two metals,
the normal Green's function in the barrier is
\begin{equation}
g_{\sigma}={{\omega_n\cos\chi/2+i\sqrt{\omega_n^2+\Delta_0^2} \sin\chi/2}\over
{\sqrt{\omega_n^2+\Delta_0^2} \cos\chi/2+i \omega_n\sin\chi/2}} ,
\label{g}
\end{equation}
where
\begin{equation}
\chi = {\phi}\mp {Z\over\cos\varphi}
 - {2i\omega_n d\over \hbar{\rm v}_0\cos\varphi}\
 \label{HI}
\end{equation}
for $\sigma=\uparrow,\downarrow$ respectively, and
\begin{equation}
Z={2hd\over \hbar {\rm v}_0}
\label{Z}
\end{equation}
is the parameter measuring the ferromagnetic barrier influence,
\noindent
$\varphi$ being the angle between the direction of ${\rm v}_0$
and the axis perpendicular to the barrier.
For $Z=0$, the Green's function, Eq.  (2.2), has  the well known
form for S/N/S junction.\cite{svi}

For spherical  Fermi surface,  the Josephson current
through the barrier of area $S$, $I=jS$,  is
\begin{equation}
I ={\pi k_{\rm B}T\over 2eR_{\rm N}}\sum_{\omega_n,\sigma}
 \int^\infty_1 {{\rm Im} [g_{\sigma}(u)
+g_{\sigma}(-u)]\over
u^3 }\,{\rm d}u,
\label{I}
\end{equation}
where $u=1/\cos \varphi$, and
the normal resistance is given by $R_{\rm N}^{-1}=e^2{\rm v}_0N(0)S$.
The temperature dependence of $\Delta_0=\Delta_0(T)$ is approximated by
\begin{equation}
\Delta _0 (T)=\Delta _0 (0)\tanh\left(1.74\sqrt {T_{\rm c}/T -1}\right),
\label{deltat}
\end{equation}
where $T_{\rm c}$ is the critical temperature of
superconducting electrodes.

The current-phase relation $I(\phi)$ is calculated numerically as a
function of $\phi$,  $T$ and
of parameters measuring the influence of the ferromagnetic barrier,
$Z$ and $d$. The results, illustrated in Figs. 1 and 2
for fixed thickness of the barrier, $d= 0.25 \,\xi_0(0)$,
 where $\xi_0(0)= \hbar{\rm
 v}_0/\pi\Delta_0(0)$ is the BCS coherence length at zero temperature,
  remain practically  unchanged for any  barrier thickness
 $d\sim 0.1\, \xi_0(0)$ at fixed $Z$.

 The ground state phase difference $\phi_{\rm gs}$  is determined from
 the  minimum of the
S/F/S junction energy\cite{bp}
\begin{equation}
 W(\phi)={\Phi_0 \over 2\pi}\int_0^\phi  I(\phi')\,{\rm  d}\phi',
\label{W}
\end{equation}
where $\Phi_0$ is the flux quantum,  Fig. 3.

At high temperature, $T/T_{\rm c}> 0.9$,
the current-phase relation is quite close to
\begin{equation}
I=\pm I_{\rm c}\sin \phi,
\label{J}
\end{equation}
where $\pm $ signs correspond to $0$ and $\pi$ junction, respectively.
The amplitude of oscillations in $I(\phi)$ decreases when $Z$ increases
from zero to the critical value $Z_{\rm c}\approx 1$. For $Z>Z_{\rm c}$,
the supercurrent flow in the opposite direction (Fig. 1(a)), and
$I_{\rm c}$ starts to increase. This is the signature of the
$\pi$ junction with ground state phase difference  $\phi_{\rm gs}=\pi$,
Fig. 3(a).

At low temperature,  for clean and transparent normal metal
barrier, the Andreev reflection is dominant, and deviations from
the  sinusoidal $I(\phi)$ law are pronounced.\cite{svi,lih} In the
S/F/S case, due to the influence of the exchange interaction in
the barrier, the current-phase relation is even more anharmonic.
Positive for small $Z$, where $0$ state is stable, the sign of
$I(\phi)$ for $\phi\to 0$ is negative for large $Z$, where $\pi$
state is stable, and multinode  oscillations in  $I(\phi)$  appear
in the vicinity of $Z_{\rm c}$, Fig. 1(b). In the latter case,
there are stationary states
 $I(\phi)=0$, $I'(\phi)>0$  both for $\phi=0$ ($2\pi$)  and  $\phi=\pi$. One of these
  states is stable, and the other is metastable.  For example,
at $T/T_{\rm c}=0.1$  for
$Z=0.9$ the metastable state is  $\phi=\pi$, $\phi=0$ being the
stable state, and {\it vice versa } for $Z=1.1$,  Fig. 3(b).  Zero and $\pi$
states are equally stable at $Z=Z_{\rm c}$. While $Z_{\rm c}$ is
practically temperature and barrier thickness independent, the interval of
the coexistence of $0$ and $\pi$ states, $Z_{\rm c^-}\leq Z \leq Z_{\rm c^+}$,
 rapidly decreases with increasing
 temperature.
 For $d/\xi_0(0)=0.25$, $Z_{\rm c}=1.03$
and the coexistence interval is $0.1\leq Z\leq 1.5$   at $T=0$, while
$Z_{\rm c}=1.05$
and the coexistence interval is $1.02\leq Z\leq 1.08$
   at $T/T_c=0.9$.

The appearence of the $\pi$ junction for $1<Z<4$
is  also manifested in the nonmonotonous  dependence of the critical
current $I_{\rm c}={\rm max}\vert I(\phi)\vert$
on $Z$ at given $T$, Fig. 2.   For $Z>4$, $\phi_{\rm
gs}=0$ again, and near $Z\approx 4$ metastable states would appear as
well. However, large $Z$ would induce the decoupling of S
electrodes.\cite{kup} For fixed $h$, such a nonmonotonic variation of
the critical current as a function of the barrier thickness $d$ has
been observed in Nb/Al/Gd/Al/Nb junctions.\cite{olivier}
The same effect is also predicted for the
atomic-scale S/F superlattices.\cite{vesna}

Recently measured $I(\phi)$ in $^3$He weak links\cite{mar} are of
the same form as the curves in Fig. 1(b) for $Z<1$, corresponding
to the metastable $\pi$ state. In $d$-wave S/F/S junctions, stable
and metastable states may also coexist due to the exchange
interaction. However, in this case the ground state
phase differences can vary from $0$ to $\pi$, depending not only on
$Z$ and $T$, but also on the crystal orientation of the
superconducting electrodes.\cite{rad,tanJ}

\section{Single-junction loop}

When  S/F/S junction  is a part of a superconducting loop (rf SQUID),
the ground state is determined by minimizing the SQUID  free energy
\begin{equation}
W(\phi)={\Phi_0 \over 2\pi}\left[\int_0^\phi I(\phi')\,{\rm d}\phi' +
{\Phi_0\over 4\pi L}(\phi-\phi_{\rm e})^2\right] ,
\label{WL}
\end{equation}
where $\phi_{\rm e}=2\pi \Phi_{\rm e}/\Phi_0$ represents the
normalized external magnetic flux through the loop, and $L$ is the
inductance.\cite{bp}
Therefore, the total magnetic flux through the
loop $\Phi=(\Phi_0 /2\pi)\phi$ is related to the external flux
$\Phi_{\rm e}$ by
\begin{equation}
 LI(\Phi)+ \Phi -\Phi_{\rm e}=0.
\label{GSL1}
\end{equation}
For $\Phi_{\rm e}=0$, and for $Z$ sufficiently large ($\pi$
junction), there is a flow of spontaneous supercurrent,\cite{rad}
with nonmonotonic dependence on the normalized inductance
\begin{equation}
l={2\pi\over\Phi_0} LI_{\rm c}.
\label{IND}
\end{equation}
Note that when $n\leq {\Phi_{\rm e}/\Phi_0}\leq {n+1}$, where
$n=\pm 1,\pm 2,...$, $\Phi /\Phi_0\to \Phi /\Phi_0 + n.$

From Eqs.~(\ref{I}) and (\ref{GSL1})
$\Phi(\Phi_{\rm e})$ is calculated numerically.
The results are illustrated in Figs. 4 and 5
for several values of $l$ and $Z$, for
$d=0.25\,\xi_0 (0)$.

At high temperature, $T/T_{\rm c}>0.9$, for  $l> 1$ and  $Z<Z_c$   the
double-well SQUID energy gives stable and metastable
solutions of Eq. (\ref{GSL1}), equally stable at
$\Phi_{\rm e}/\Phi_0=0.5$. Corresponding flux jump between
$\Phi/\Phi_0\approx 0$ and $\Phi/\Phi_0\approx 1$ is indicated by
vertical line  linking the   stable
solutions, Fig.~4(a). In experiments, the hysteretic loops
enclosing these lines are usually seen. For $\pi$ junctions,
$Z>Z_c$,  Fig.~4(b), there is only one stable solution with
$\Phi/\Phi_0\approx 0.5$, and flux jumps at
$\Phi_{\rm e}/\Phi_0=0$ and $1$, without hysteresis. For $l\ll 1$,
the effect of  supercurrents in the loop is
negligible, and $\Phi\approx \Phi_{\rm e}$ for any $Z$.

At low temperature,  similar results are obtained only in the
limiting cases of normal metal and strong ferromagnet barriers.
For $Z=0$ and $l\gg 1$, there is the usual integer  flux
quantization,  Fig.~5(a).  For $Z=3$ and  $l\gg 1$, there is the
half-integer flux quantization, Fig.~5(d),  $\Phi(\Phi_{\rm e})$
being nonhysteretic in contrast to the previous case.
Corresponding free energies are illustrated in Fig.~6.
New features appear  for $Z$ close to $Z_{\rm c}$, related
to the presence of stable and metastable minima of the triple-well
SQUID energy, Figs. 6(b) and 6(c).
For relatively  small inductance, $l<10$, the SQUID
has two degenerate configurations: the nearly zero-- and one-half-- fluxoid
state, when $\Phi _{\rm e}/\Phi_0<0.5$, and the nearly one-half-- and
one-- fluxoid states, when $\Phi _{\rm e}/\Phi_0>0.5$, as shown in
Fig. 7. For example, at
$T/T_c=0.1$, for $0.4\leq Z \leq 1.4$ and $l<10$,  new hysteretic jumps
of $\Phi$ are found, from $\Phi\approx 0$ to $\Phi/\Phi_0\approx 0.5$ in
the interval $0<\Phi_{\rm e}/\Phi_0<0.5$, and from $\Phi/\Phi_0\approx
0.5$  to $\Phi/\Phi_0 \approx 1$ in  the interval $0.5<\Phi_{\rm
e}/\Phi_0<1$, Figs.~5(b) and 5(c). This anomalous periodicity of
$\Phi(\Phi_{\rm e})$ resemble to the usual one, Fig.~5(a),
scaled by  $\Phi_0\to \Phi_0/2$.

\section{Two-junction loop}

For a loop containing two S/F/S  junctions (dc SQUID) the total phase
difference  is $\phi=\phi_1-\phi_2$,
where $\phi_1$ and $\phi_2$ are the phase differences at the first and
the second junction,
respectively. In the ground state, the magnetic flux
through the loop $\Phi(\Phi_{\rm e})$
is calculated from two coupled equations\cite{bp}
\begin{equation}
 I_1(\Phi_1)+I_2(\Phi_2)=I_{\rm e}
 \label{struje}
\end{equation}
and
\begin{equation}
 L_1I_1(\Phi_1)-L_2I_2(\Phi_2)+ \Phi-\Phi_{\rm e}=0,
 \label{GSL2}
\end{equation}
where $I_{\rm e}$ is the external current, $\Phi_{1,2}=(\Phi_0/2\pi)\phi_{1,2}$
and $\Phi=\Phi_1-\Phi_2$. As expected, $\Phi(\Phi_{\rm e})$ dependence
for dc SQUID in the absence of the external current, $I_{\rm e}=0$, is
similar to that for   rf SQUID.

At high temperature, $T/T_{\rm c}>0.9$, and
for two  $0$ junctions, $Z_1<Z_{\rm c}$, $Z_2<Z_{\rm c}$,
or two $\pi$ junctions, $Z_1>Z_{\rm c}$, $Z_2>Z_{\rm c}$,
$\Phi(\Phi_{\rm e})$
dependence is qualitatively the same as in a hysteretic
single $0$ junction loop, Fig.~4(a). For one $0$ junction
and one $\pi$ junction, the two-junction loop behaves as
a nonhysteretic single $\pi$ junction loop, Fig.~4(b).

At low temperature, outside  the coexistence  interval of $0$ and $\pi$
states, for two $0$ junctions, $Z_1<Z_{\rm c^-}$, $Z_2<Z_{\rm c^-}$,
or for two $\pi$ junctions, $Z_1>Z_{\rm c^+}$, $Z_2>Z_{\rm c^+}$, the
two-junction loop behaves as a hysteretic single $0$ junction loop. For
one $0$ junction, and one $\pi$ junction, $Z_1<Z_{\rm c^-}$, $Z_2>Z_{\rm
c^+}$ or {\it vice versa}, the two-junction loop behaves as a
nonhysteretic single $\pi$ junction loop. For example, at $T/T_c=0.1$,
$\Phi(\Phi_{\rm e})$ curves for $Z_1=Z_2=0$ and $Z_1=Z_2=3$ would be
practically the  same as in Fig.~5(a), and for $Z_1=0$ and $Z_2=3$ as
in Fig.~5(d), if $L_1=L_2$, the first junction normalized inductance
$l_1=(2\pi/\Phi_0)I_{\rm c1}L_1$ corresponding to $l$ of a
single-junction loop. When  $Z$, for at least one of the junctions,
lies within the coexistence interval, $Z_{\rm c^-}<Z< Z_{\rm c^+}$ new
jumps between the stable solutions appear for sufficiently small
inductance, similarly to the single-junction loop. For example, at
$T/T_c=0.1$ and for $L_1=L_2$, $\Phi(\Phi_{\rm e})$ curves for $Z_1=0$
and $Z_2=0.9$ are the same as in Fig.~5(b), and for $Z_1=0$ and
$Z_2=1.1$, as in Fig.~5(c), see Fig.~8. Magnetic flux penetration in the case when
metastable states exist in both junctions is illustrated for $Z_1=1.1$
and $Z_2=0.9$ in Fig.~9(a).

For nonzero external current, up to  its maximum value $I_{\rm
ec}(\Phi_e)\leq I_{\rm c1}+I_{\rm c2}$, $\Phi (\Phi_{\rm e})$
curves are similar to those for $I_{\rm e }=0$, the flux through
the loop becoming closer to the external flux. However, when $0$
and $\pi$ states coexist, for high external current  flux jumps appear
 for any inductance, as illustrated for $I_{\rm e}=I_{\rm ec}(\Phi_{\rm
e})$ in Fig.~9(b).

In general,  $I_{\rm ec}(\Phi_{\rm e})$ curves have characteristic dips
at the same positions as the jumps in $\Phi (\Phi_{\rm e})$ curves.
Three typical cases  at low temperature are illustrated in   Fig.~10 for
one period of $I_{\rm ec}(\Phi_{\rm e})$. For two $0$ junction loop,
there is the usual $I_{\rm ec}(\Phi_e)$ dependence,\cite{bp}  with a
single dip at $\Phi_{\rm e}/\Phi_0=0.5$, Fig.~10(a). For one $0$   and
one $\pi$ junction, there are two dips at the ends of the period,
$\Phi_{\rm e}/\Phi_0=0$ and $1$, Fig. 10(b). New characteristic behavior
is obtained when there is the coexistence of $0$  and $\pi$ states  in
one of the junctions at least: $I_{\rm ec}(\Phi_e)$ curve  has three
maxima and two central dips at the flux jumps positions in $\Phi
(\Phi_{\rm e})$, Fig.~10(c).

\section{Summary and discussion}

The ground state phase difference $\phi_{\rm gs}$ in $s$-wave S/F/S
junctions is always $0$ or $\pi$. In the vicinity of the critical
strength of the ferromagnetic barrier influence $Z _{\rm c}$, where $0$
and $\pi$ states are equally stable, the characteristic multinode
anharmonicity  of $I(\phi)$ implies the coexistence of stable $0$  and
metastable $\pi$  state for $Z_{\rm c^-}\leq Z \leq Z_{\rm c}$, or {\it
vice versa} for $Z_{\rm c}\leq Z\leq Z_{\rm c^+}$.
For $Z<Z_{\rm c^-}$ and for $Z>Z_{\rm c^+} $ only stable states exist,
$\phi_{\rm gs}=0$, and $\phi_{\rm gs}=\pi$, respectively. The interval of
$Z$, where the metastable states appear, rapidly decreases with
increasing temperature, whereas $Z_{\rm c}$ is practically unchanged.

Stable and metastable states can also coexist in $d$-wave S/F/S
junctions,\cite{rad} in contrast to $d$-wave non-magnetic junctions
where the similar anharmonicity of the current-phase relation does
not lead to the metastable states.\cite{tan97}

The peculiar low temperature behavior of clean and transparent S/F/S
junctions is due to the modification of the Andreev reflection by the
exchange interaction in the ferromagnet. The effects of the barrier
interfacial nontransparency, and of the scattering on impurities,
neglected in this paper, suppress the Andreev reflection. As a
consequence, anharmonicity of the current-phase relation becomes less
pronounced, the value of $Z_{\rm c}$ changes, and the metastability region
narrows.\cite{zor} However, when these effects are relatively weak, they
do not change the above results qualitatively. We also expect that the
suppression of the pair potential near the interfaces, not included in
the present non-selfconsistent calculation, would not destroy the
metastable states neither.\cite{tanJ}

The coexistence of $0$ and $\pi$  states in the S/F/S-junction leads to
the coexistence of integer and half-integer fluxoid configurations in
SQUIDs, corresponding to the minima of the triple-well potential
energy, and generating two flux jumps per one external flux quantum.
This is the case when $Z$ is close to $Z_{\rm c}$ in rf SQUID, and at
least in one of the junctions in dc SQUID. Another feature  is  the
multipeak dependence of the dc SQUID critical current on the external
magnetic flux.

In conclusion, experimental investigation of the predicted
new manifestation of the modified Andreev reflection
is desirable. Beside possible device applications, e.g. quantum
computing,\cite{orlando} the coexistence of integer and half-integer
flux quantum  configuration in S/F/S junction SQUIDs could be also used
for the experimental study of the quantum superposition of
macroscopically distinct states.\cite{leg}

\newpage

\begin{figure}
\caption{\label{Fig}
Current-phase relation of S/F/S junction
$I(\phi)$ for several values of $Z$, including $Z_{\rm c}$ (dashed
curve), for $d/\xi_0 (0)=0.25$. (a) $T/T_{\rm c}=0.9$, $Z_{\rm
c}=1.049$, and (b) $T/T_{\rm c}=0.1$, $Z_{\rm c}=1.030$. }
\end{figure}

\begin{figure}
\caption{\label{Fig2}
Critical current $I_{\rm c}$ as a function
of $Z$ for  $T/T_{\rm c}=0.1$ and $0.9$.}
\end{figure}

\begin{figure}
\caption{\label{Fig3} Free energy of S/F/S junction $W$ as a
function of $\phi$ for several values of $Z$, including $Z_{\rm
c}$ (dashed curves), for (a) $T/T_{\rm c}=0.9$  and (b) $T/T_{\rm
c}=0.1$.}
\end{figure}

\begin{figure}
\caption{\label{Fig4} Magnetic flux $\Phi$ through the
single-junction loop as a function of the external flux $\Phi_{\rm
e}$ at high temperatures, $T/T_{\rm c}> 0.9$, for three values of
the normalized inductance $l$. (a) $Z<Z_{\rm c},$ and  (b)
$Z>Z_{\rm c}$. Only stable (solid curves) and metastable
(dashed curves) solutions are shown, vertical lines  linking the stable solutions.
}
\end{figure}

\begin{figure}
\caption{\label{Fig5} Magnetic flux $\Phi$  through the
single-junction loop as a function of the external flux $\Phi_{\rm
e}$ at low temperature, $T/T_{\rm c}=0.1$, for three values of the
normalized inductance $l$. (a) $Z=0$, (b) $Z=0.9$, (c) $Z=1.1$,
and (d) $Z=3$. Only stable  (solid curves) and metastable
(dashed curves) solutions are shown.
Vertical lines are drawn  to link the stable solutions.
}
\end{figure}

\begin{figure}
\caption{Free energy of the single-junction loop $W$ as a function
of $\Phi$ at low temperature, $T/T_{\rm c}=0.1$, for $l^{-1}=0.1$,
four values of $Z$, and for $\Phi_{\rm e}/\Phi_0=0$ (long-dashed
curves), $0.5$ (solid curves), and $1$ (short-dashed curves).}
\label{Fig6}
\end{figure}

\begin{figure}
\caption{Free energy of the single-junction loop $W$ as a function
of $\Phi$ at low temperature, $T/T_{\rm c}=0.1$, for $l^{-1}=0.5$,
and for $\Phi{\rm e}/\Phi_0$ corresponding to the flux jumps at
Figs. 5(b) and 5(c): (a) $Z=0.9$, $\Phi_{\rm e}/\Phi_0=0.37$
(solid curve), $\Phi_{\rm e}/\Phi_0=0.63$ (dashed curve); (b)
$Z=1.1$, $\Phi_{\rm e}/\Phi_0=0.20$ (solid curve), $\Phi_{\rm
e}/\Phi_0=0.80$ (dashed curve). } \label{Fig7}
\end{figure}

\begin{figure}
\caption{Magnetic flux $\Phi$  through the two-junction loop as a
function of the external flux $\Phi_{\rm e}$ at low temperature,
$T/T_{\rm c}=0.1$, for $L_1=L_2$, several values of $Z_1$ and
$Z_2$, and for two values of the first junction  normalized
inductance $l_1$ in the absence of the external current, $I_{\rm
e}=0$. } \label{Fig8}
\end{figure}

\begin{figure}
\caption{\label{Fig9}
Magnetic flux $\Phi$  through
the two-junction loop, $Z_1=1.1$, $Z_2=0.9$, $L_1=L_2$,
as a function of the external flux $\Phi_{\rm e}$
at low temperature, $T/T_{\rm c}=0.1$, for two values of the
first junction  normalized inductance $l_1$. (a) $I_{\rm e}=0$ and (b) $I_{\rm e}=I_{\rm
ec}$.
}
\end{figure}

\begin{figure}
\caption{Maximum external current $I_{\rm ec}$ of the two-junction loop as
a function of the external flux $\Phi_{\rm e}$
at low temperature, $T/T_{\rm c}=0.1$, for two values of the
first junction  normalized inductance $l_1$. Three typical cases are shown:
(a) $Z_1=0$, $Z_2=0$, (b) $Z_1=0$,$Z_2=3$, (c) $Z_1=1.1$,$Z_2=0.9$.
}
\label{Fig10}
\end{figure}

\end{document}